\documentclass[smallcondensed]{svjour3}
\usepackage{epsfig}
\usepackage{amsmath}
\usepackage{bm}
\usepackage{epsfig}
\usepackage{epstopdf}
\usepackage{subfig}
\usepackage{graphicx}
\usepackage[section]{placeins}
\usepackage[toc]{appendix}
\def\be{\begin{equation}}
\def\ee{\end{equation}}
\def\bea{\begin{eqnarray}}
\def\eea{\end{eqnarray}}

\begin{document}

\title{Temporal dynamics and nonclassical photon statistics of quadratically
coupled optomechanical systems}

\author{Shailendra Kumar Singh \and S. V. Muniandy}
\institute{Corresponding Author: Shailendra Kumar Singh\at
              Institute of Nuclear Sciences, Hacettepe University, 06800, Ankara, Turkey\\
              \email{singhshailendra3@gmail.com} 
                                  \and S.V. Muniandy
            \at Center of Theoretical and Computational Physics, Department of Physics, University of Malaya, 50603 KualaLumpur, Malaysia\\
                            \email{msithi.um.edu.my}}
            
\maketitle              
\begin{abstract}
Quantum Optomechanical system serves as an interface for coupling between
photons and phonons due to mechanical oscillations. We use the
Heisenberg-Langevin approach under Markovian white noise approximation to
study a quadratically coupled optomechanical system which contains a thin
dielectric membrane quadratically coupled to the cavity field. A
decorrelation method is employed to solve for a larger number of coupled
equations. Transient mean number of cavity photons and phonons that provide
dynamical behaviour are computed under different coupling regime. We have
also obtained the two-boson second-order correlation functions for the cavity
field, membrane oscillator and their cross correlations that provide
nonclassical properties governed by quadratic optomechanical system.
\end{abstract}

\section{\qquad \protect\bigskip INTRODUCTION}

Recent years have seen significant experimental progress in realizing
deterministic interactions between single photons, which has profound
importance for future optical technologies. Most novel experiments have been
explored in cavity quantum electrodynamics (cQED) \cite{Haroche,Walther},
where photons inherent the saturation of a single two-level atom due to
strong interactions between the atom and the cavity field. A single
atom-cavity system described by well-known Jaynes-Cummings model \cite{Jaynes}
has been used as an important test bed for implementation of quantum
information algorithms as well as construction of a quantum network with the
aim for quantum computation. The strong-coupling regime of the cQED is
reached when the coupling strength of the atom with the cavity mode
dominates over the decoherence processes, due to spontaneous emission from
the excited atomic level to ground state and the leakage of photons in the
cavity mode. Based on Fabry- Perot interferometry \cite{Yablonovitch},
semiconductor microcavities have been developed where excitons act as
quantum systems \cite{Gerard}. Alternative approaches have been explored
based on slow-light enhanced Kerr nonlinearites \cite{Lin}, single
dye-molecules \cite{Hwang}, strong photon interactions mediated by Rydberg
atoms \cite{Dudin}, atoms coupled to hollow-core fibers \cite{Saha} and
nanofiber traps \cite{Kolchin}.\newline

Optomechanical system provides a more promising approach towards realizing
strong photon interactions \cite{Kippenberg}. In comparison to atomic
systems, for example macroscopic mechanical resonators are relatively easy
to control and to be integrated with other systems. In these systems, an
optical cavity, with a movable end mirror \cite{Grob,Wilson} or with a
micromechanical membrane is subjected to mechanical effect caused by light
through radiation pressure \cite{Thomp}. So, Cavity Quantum Optomechanics
has emerged as an interesting area for studying quantum features at the
mesoscale, where it is possible to control the quantum state of mechanical
oscillators by their coupling to light field \cite{Mas}. Recent advances in
this area includes the realization of quantum-coherent coupling of a
mechanical oscillator with an optical cavity \cite{Verhagen}, where the
coupling rate exceeds both the cavity and mechanical motion decoherence
rate, laser cooling of a nanomechanical oscillator to its ground state \cite {Chan}.\newline

Experimental and theoretical proposals aiming to study quantum aspects of
interactions between the optical cavities and mechanical objects have
focused on cavities in which one of the cavity's mirror is free to move for
example, in response to radiation pressure exerted by light in the cavity.
This nearly includes all optomechanical systems described in the literature,
including cavities with 'folded' geometries, cavities in which multiple
mirrors are free to move \cite{Bose}, and whispering gallery mode resonators 
\cite{Santamore} in which light is confined to a waveguide. All these
methods have two important considerations. First, cavity's detuning is
proportional to the displacement of a mechanical degree of freedom, which is
mirror displacement or waveguide elongation. Second, a single device must
provide both optical confinement and mechanical effects. So, achieving good
mechanical and optical properties both simultaneously has been the main aim
in quantum optomechanical systems because high-finesse mirrors are not
easily integrated into micromachined devices. Experimental work done so far
have achieved sufficient optomechanical coupling to laser-cooled mechanical
devices \cite{Hohberger,Arcizet} but have not yet reached to quantum regime
due to technical problems mentioned above. A more fundamental challenge is
to read out the mechanical element's energy eigenstate. Displacement
measurements cannot determine an oscillator's energy eigenstate and
measurements coupling to quantities other than displacement have been
difficult to realize in practice \cite{Martin,Jacobs}.\newline

In this work, we theoretically study an optomechanical system realized in
seminal experimental works \cite{Zwickl,Zwickll} and has potential to
resolve both of these above-mentioned challenges. We study a system which
contains a thin dielectric membrane with quadratic response to the cavity
fields. Coupled Heisenberg-Langevin equations are obtained under Markovian
white noise approximation, and solutions up to second-order correlation
operators have been developed in Section II. In the same section, we also
discuss the transient dynamics of the system Hamiltonian in various
scenario. Section III discusses the results with the nonclassical photon
statistics of the cavity mode as well as mechanical oscillations mode
including cross-correlation between them. We conclude our results in Section
IV.

\section{\protect\bigskip Heisenberg-Langevin Approach For Optomechanics}

We consider a quadratically coupled optomechanical resonator formed by a
micropillar with moveable Bragg reflectors and a thin dielectric membrane at
the node (or antinode) of the resonator. The mechanical displacement of the
membrane quadratically couples to the cavity photon number. In addition, we
have a monochromatic laser field with frequency $\omega _{L}$ applied to
drive the cavity. The corresponding Hamiltonian for the scheme is given by, 
\begin{equation}
\hat{H}=\hat{H}_{0}+\hat{H}_{R},
\end{equation}%
where $\hat{H}_{0}$ and $\hat{H}_{R}$ are given by Eqs. (2) and (3). In our
model, $\hat{H}_{R}$ is the part where all the two subsystem cavity photons,
phonons (due to mechanical motion of the membrane) respectively interact
with their corresponding reservoirs which are collection of harmonic
oscillators. We write 
\begin{equation}
\hat{H}_{0}=\omega _{c}\hat{a}^{\dagger }\hat{a}+\omega _{M}\hat{b}^{\dagger
}\hat{b}+g_{opt}\left[ \hat{a}^{\dagger }\hat{a}\left( \hat{b}^{\dagger }+%
\hat{b}\right) ^{2}\right] +\Omega \left[ e^{-i\omega _{L}t}\hat{a}^{\dagger
}\text{ }+e^{i\omega _{L}t}\hat{a}\right]
\end{equation}
where $\hat{a}$ \ and $\hat{b}$ \ correspond to field operators for cavity
photons and\ phonons (due to mechanical motion of the membrane) with
frequencies $\omega _{c}$ and $\omega _{M}$, respectively. The third term in
R.H.S of Eq. (2) describes the quadratic optomechanical coupling with
strength $g_{opt}$ between the cavity field and the mechanical motion of the
membrane and the fourth term describes driving process of the cavity field,
with $\Omega $ is the Rabi frequency of the driving field. Meanwhile, $\hat{H%
}_{R}$ takes the following form: 
\begin{equation}
\hat{H}_{R}=\text{ }\underset{}{\overbrace{\overset{\text{for cavity mode}}{%
\sum_{k}\omega _{k}\hat{m}_{k}^{\dagger }\hat{m}_{k}+\sum_{k}g_{k}\left( 
\hat{m}_{k}^{\dagger }\hat{a}+\hat{a}^{\dagger }\hat{m}_{k}\right) }}}+%
\overbrace{\overset{\text{for moving membrane}}{\sum_{k^{\prime }}\omega
_{k^{\prime }}\hat{n}_{k^{\prime }}^{\dagger }\hat{n}_{k^{\prime
}}+\sum_{k^{\prime }}g_{k^{\prime }}\left( \hat{n}_{k^{\prime }}^{\dagger }%
\hat{b}+\hat{b}^{\dagger }\hat{n}_{k^{\prime }}\right) }},
\end{equation}%
where the first two terms on the R.H.S of Eq.(3) represent the damping of
cavity mode through reservoir of harmonic oscillator operators $\hat{m}_{k}$
and $\hat{m}_{k}^{\dagger }$; while the third and fourth terms are responsible
for damping of membrane through the harmonic oscillator operators $\hat{n}_{k^{\prime
}}$ and $\left( \hat{n}_{k^{\prime }}^{\dagger }\right) $. In the rotating
frame with the driving frequency $\omega _{L}$, the Hamiltonian $\hat{H}_{0}$
reduces to 
\begin{equation}
\hat{H}_{0}=\Delta _{c}\hat{a}^{\dagger }\hat{a}+\omega _{M}\hat{b}^{\dagger
}\hat{b}+g_{opt}\left[ \hat{a}^{\dagger }\hat{a}\left( \hat{b}^{\dagger }+%
\hat{b}\right) ^{2}\right] +\Omega \left( \hat{a}^{\dagger }\text{ }+\hat{a}%
\right),
\end{equation}%
where $\Delta _{c}=\left( \omega _{c}-\omega _{L}\right) $ is the
corresponding detuning from the driving laser frequency $\omega _{L}.$
Expanding the quadratic term \ $g_{opt}\left[ \hat{a}^{\dagger }\hat{a}%
\left( \hat{b}^{\dagger }+\hat{b}\right) ^{2}\right] ,$ we have the
Hamiltonian $\hat{H}_{0}$\ in the following form: 
\begin{equation}
\hat{H}_{0}=\Delta _{c}\hat{a}^{\dagger }\hat{a}+\omega _{M}\hat{b}^{\dagger
}\hat{b}+g_{opt}\left[ \hat{a}^{\dagger }\hat{a}\left( \hat{b}^{\dagger 2}+%
\hat{b}^{2}+2\hat{b}^{\dagger }\hat{b}+1\right) \right] +\Omega \left( \hat{a%
}^{\dagger }\text{ }+\hat{a}\right)
\end{equation}
\newline
Coupling strength $g_{opt}$ should satisfy the condition $\left( \omega
_{M}+4sg_{opt}\right) >0$ for the stability of the membrane \cite{Nori},
where $s$ is the number of photons inside the cavity. The corresponding
Heisenberg equations of motion for the field operators of different
subsystem (cavity, membrane) are given by Eqs. (6) and (8), whereas the
corresponding equations of motion for reservoir operators are given by Eqs.
(7) and (9): 
\begin{equation}
\frac{d}{dt}\hat{a}=-i\Delta _{c}\hat{a}-ig_{opt}\left[ \hat{a}\left( \hat{b}%
^{\dagger 2}+\hat{b}^{2}+2\hat{b}^{\dagger }\hat{b}+1\right) \right]
-i\Omega -i\sum_{k}g_{k}\hat{m}_{k}.
\end{equation}%
\begin{equation}
\frac{d}{dt}\hat{m}_{k}=-i\omega _{k}\hat{m}_{k}-ig_{k}\hat{a}.
\end{equation}%
\begin{equation}
\frac{d}{dt}\hat{b}=-i\omega _{M}\hat{b}-2ig_{opt}\left[ \hat{a}^{\dagger }%
\hat{a}\left( \text{ }\hat{b}+\hat{b}^{\dagger }\right) \right]
-i\sum_{k^{\prime }}g_{k^{\prime }}\hat{n}_{k^{\prime }}.
\end{equation}%
\begin{equation}
\frac{d}{dt}\hat{n}_{k^{\prime }}=-i\omega _{k^{\prime }}\hat{n}_{k^{\prime
}}-ig_{k^{\prime }}\hat{b}.
\end{equation}%
Now the corresponding operator equations for different reservoir need to be
explicitly solved further to remove the terms of reservoir operators in Eqs
(6) and (8). Integrating Eq. (7), one gets 
\begin{equation}
\hat{m}_{k}(t)=\hat{m}_{k}(0)e^{-i\omega
_{k}t}-ig_{k}\int_{0}^{t}dt^{^{\prime }}\hat{a}(t^{^{\prime }})e^{-i\omega
_{k}(t-t^{^{\prime }})}.
\end{equation}%
In Eq. (10), the first term represents the free evolution of the reservoir
modes, whereas the second term arises from their interaction with the
cavity. The reservoir modes $\hat{m}_{k}(t)$ can be eliminated from the Eq.
(6) by substituting the value of $\hat{m}_{k}(t)$ from Eq. (10) in Eq. (6).
So, we have 
\begin{equation}
\frac{d}{dt}\hat{a}=-i\Delta _{c}\hat{a}-ig_{opt}\left[ \hat{a}\left( \hat{b}%
^{\dagger 2}+\hat{b}^{2}+2\hat{b}^{\dagger }\hat{b}+1\right) \right]
-i\Omega -\sum_{k}g_{k}^{2}\int_{0}^{t}dt^{^{\prime }}\hat{a}(t^{^{\prime
}})e^{-i\omega _{k}(t-t^{^{\prime }})}+F_{a}(t),
\end{equation}%
where $F_{a}(t)=-i\sum_{k}g_{k}\hat{m}_{k}(0)e^{-i\omega _{k}t}$ \ is the
noise operator for the cavity which depends on the reservoir variables. \
The term containing the integral is expressed as 
\begin{equation}
\sum_{k}g_{k}^{2}\int_{0}^{t}dt^{^{\prime }}\hat{a}(t^{^{\prime
}})e^{-i\omega _{k}(t-t^{^{\prime }})}\simeq \frac{1}{2}\Gamma _{a}\hat{a}%
(t).
\end{equation}%
Thus, from Eqs. (11) and (12), we obtain 
\begin{equation}
\frac{d}{dt}\hat{a}=-i\Delta _{c}\hat{a}-ig_{opt}\left[ \hat{a}\left( \hat{b}%
^{\dagger 2}+\hat{b}^{2}+2\hat{b}^{\dagger }\hat{b}+1\right) \right]
-i\Omega -\frac{1}{2}\Gamma _{a}\text{ }\hat{a}(t)+F_{a}(t),
\end{equation}%
where $\Gamma _{a}$ is the damping constant for cavity mode. Similarly, we have equation for mechanical motion
of the membrane as follows: 
\begin{equation}
\frac{d}{dt}\hat{b}=-i\omega _{M}\hat{b}-2ig_{opt}\left[ \hat{a}^{\dagger }%
\hat{a}\left( \hat{b}+\hat{b}^{\dagger }\right) \right] -\frac{1}{2}\Gamma
_{b}\text{ }\hat{b}+F_{b}(t);
\end{equation}%
where $F_{b}(t)$ and $\Gamma _{b}$ are noise operator and damping constant
for moving membrane respectively, given by 
\begin{equation}
F_{b}(t)=-i\sum_{k^{\prime }}g_{k^{\prime }}\hat{n}_{k^{\prime
}}(0)e^{-i\omega _{k^{\prime }}t},\text{ }
\end{equation}
and 
\begin{equation}
\sum_{k^{\prime }}g_{k^{\prime }}^{2}\int_{0}^{t}dt^{^{\prime }}\hat{b}%
(t^{^{\prime }})e^{-i\omega _{k^{\prime }}(t-t^{^{\prime }})}\simeq \frac{1}{%
2}\Gamma _{b}\hat{b}(t).
\end{equation}

\subsection{Quantum Correlations}

We have obtained the coupled equations involving the mean of the operators $%
\hat{a}$ and $\hat{b}$ ,\ their corresponding adjoints and their odd and
even products. Since both the reservoirs are big as well as always in
thermal equilibrium, we have $\ \left\langle F_{a}(t)\right\rangle =$ $%
\left\langle F_{b}(t)\right\rangle =0$. We also have, $\left\langle F_{\hat{a%
}^{\dagger }}(t)\hat{b}(t)\right\rangle =\left\langle F_{\hat{b}}(t)\text{ }%
\hat{a}(t)\right\rangle =\left\langle F_{a}(t)\hat{b}^{\dagger
}(t)\right\rangle =\left\langle F_{\hat{b}^{\dagger }}(t)\text{ }\hat{a}%
(t)\right\rangle =0.$ \ Similarly, quantum correlations between any
subsystem (cavity, moving membrane)\ with different noise operator other
than its own coupled reservoir's noise operator vanish altogether. The
following set of equations has been obtained by using the
Heisenberg-Langevin approach given in \cite{Scully}. 
\begin{eqnarray}
\frac{d}{dt}\left\langle \hat{a}\right\rangle &=&-i\Delta _{c}\left\langle 
\hat{a}\right\rangle -ig_{opt}\left[ \left\langle \hat{a}\hat{b}^{\dagger
2}\right\rangle +\left\langle \hat{a}\hat{b}^{2}\right\rangle +2\left\langle 
\hat{a}\hat{b}^{\dagger }\hat{b}\right\rangle +\left\langle \hat{a}%
\right\rangle \right]  \notag \\
&&-i\Omega -\frac{1}{2}\Gamma _{a}\left\langle \hat{a}\right\rangle .
\end{eqnarray}%
\begin{eqnarray}
\frac{d}{dt}\left\langle \hat{a}^{\dagger }\right\rangle &=&i\Delta
_{c}\left\langle \hat{a}^{\dagger }\right\rangle +ig_{opt}\left[
\left\langle \hat{a}^{\dagger }\hat{b}^{\dagger 2}\right\rangle
+\left\langle \hat{a}^{\dagger }\hat{b}^{2}\right\rangle +2\left\langle \hat{%
a}^{\dagger }\hat{b}^{\dagger }\hat{b}\right\rangle +\left\langle \hat{a}%
^{\dagger }\right\rangle \right]  \notag \\
&&+i\Omega -\frac{1}{2}\Gamma _{a}\left\langle \hat{a}^{\dagger
}\right\rangle.
\end{eqnarray}%
\begin{equation}
\frac{d}{dt}\left\langle \hat{b}\right\rangle =-i\omega _{M}\left\langle 
\hat{b}\right\rangle -2ig_{opt}\left[ \left\langle \hat{a}^{\dagger }\hat{a}%
\hat{b}\right\rangle +\left\langle \hat{a}^{\dagger }\hat{a}\hat{b}^{\dagger
}\right\rangle \right] -\frac{1}{2}\Gamma _{b}\left\langle \hat{b}%
\right\rangle.
\end{equation}%
\begin{equation}
\frac{d}{dt}\left\langle \hat{b}^{\dagger }\right\rangle =i\omega
_{M}\left\langle \hat{b}^{\dagger }\right\rangle +2ig_{opt}\left[
\left\langle \hat{a}^{\dagger }\hat{a}\hat{b}^{\dagger }\right\rangle
+\left\langle \hat{a}^{\dagger }\hat{a}\hat{b}\right\rangle \right] -\frac{1%
}{2}\Gamma _{b}\left\langle \hat{b}^{\dagger }\right\rangle.
\end{equation}%
\begin{equation}
\frac{d}{dt}\left\langle \hat{a}^{\dagger }\hat{a}\right\rangle =-i\Omega
\left( \left\langle \hat{a}^{\dagger }\right\rangle -\left\langle \hat{a}%
\right\rangle \right) -\Gamma _{a}\left\langle \hat{a}^{\dagger }\hat{a}%
\right\rangle +\Gamma _{a}\bar{n}_{th}^{a}.
\end{equation}%
\begin{equation}
\frac{d}{dt}\left\langle \hat{b}^{\dagger }\hat{b}\right\rangle =-2ig_{opt}%
\left[ \left\langle \hat{a}^{\dagger }\hat{a}\text{ }\hat{b}^{\dagger
2}\right\rangle -\left\langle \hat{a}^{\dagger }\hat{a}\text{ }\hat{b}%
^{2}\right\rangle \right] -\Gamma _{b}\left\langle \hat{b}^{\dagger }\hat{b}%
\right\rangle +\Gamma _{b}\bar{n}_{th}^{b}.
\end{equation}

\begin{eqnarray}
\frac{d}{dt}\left\langle \hat{a}\hat{b}^{\dagger }\right\rangle &=&i\left(
\omega _{M}-\Delta _{c}\right) \left\langle \hat{a}\hat{b}^{\dagger
}\right\rangle -i\Omega \left\langle \hat{b}^{\dagger }\right\rangle -\frac{1%
}{2}\left( \Gamma _{a}+\Gamma _{b}\right) \left\langle \hat{a}\hat{b}%
^{\dagger }\right\rangle \\
&&+ig_{opt}\left[ 2\left\langle \hat{a}^{\dagger }\hat{a}^{2}\hat{b}%
\right\rangle -\left\langle \hat{a}\hat{b}^{\dagger }\hat{b}%
^{2}\right\rangle -\left\langle \hat{a}\hat{b}^{\dagger 3}\right\rangle
+2\left( \left\langle \hat{a}^{\dagger }\hat{a}^{2}\hat{b}^{\dagger
}\right\rangle -\left\langle \hat{a}\hat{b}^{\dagger 2}\hat{b}\right\rangle
\right) -\left\langle \hat{a}\hat{b}^{\dagger }\right\rangle \right].  \notag
\end{eqnarray}

\begin{eqnarray}
\frac{d}{dt}\left\langle \hat{a}^{\dagger }\hat{b}\right\rangle &=&i\left(
\Delta _{c}-\omega _{M}\right) \left\langle \hat{a}^{\dagger }\hat{b}%
\right\rangle +i\Omega \left\langle \hat{b}\right\rangle -\frac{1}{2}\left(
\Gamma _{a}+\Gamma _{b}\right) \left\langle \hat{a}^{\dagger }\hat{b}%
\right\rangle \\
&&+ig_{opt}\left[ \left\langle \hat{a}^{\dagger }\hat{b}^{\dagger 2}\hat{b}%
\right\rangle +\left\langle \hat{a}^{\dagger }\hat{b}^{3}\right\rangle
-2\left\langle \hat{a}^{\dagger 2}\hat{a}\hat{b}^{\dagger }\right\rangle
+2\left( \left\langle \hat{a}^{\dagger }\hat{b}^{\dagger }\hat{b}%
^{2}\right\rangle -\left\langle \hat{a}^{\dagger 2}\hat{a}\hat{b}%
\right\rangle \right) +\left\langle \hat{a}^{\dagger }\hat{b}\right\rangle %
\right].  \notag
\end{eqnarray}

\begin{eqnarray}
\frac{d}{dt}\left\langle \hat{a}\hat{b}\right\rangle &=&-i\left( \Delta
_{c}+\omega _{M}\right) \left\langle \hat{a}\hat{b}\right\rangle -i\Omega
\left\langle \hat{b}\right\rangle -\frac{1}{2}\left( \Gamma _{a}+\Gamma
_{b}\right) \left\langle \hat{a}\hat{b}\right\rangle -ig_{opt}\left[
2\left\langle \hat{a}\hat{b}^{\dagger }\right\rangle +\left\langle \hat{a}%
\hat{b}^{\dagger 2}\hat{b}\right\rangle \right.  \notag \\
&&\left. +2\left\langle \hat{a}^{\dagger }\hat{a}^{2}\hat{b}^{\dagger
}\right\rangle +\left\langle \hat{a}\hat{b}^{3}\right\rangle +2\left(
\left\langle \hat{a}\hat{b}\right\rangle +\left\langle \hat{a}\hat{b}%
^{\dagger }\hat{b}^{2}\right\rangle +\left\langle \hat{a}^{\dagger }\hat{a}%
^{2}\hat{b}\right\rangle \right) +\left\langle \hat{a}\hat{b}\right\rangle %
\right].
\end{eqnarray}

\begin{eqnarray}
\frac{d}{dt}\left\langle \hat{a}^{\dagger }\hat{b}^{\dagger }\right\rangle
&=&i\left( \Delta _{c}+\omega _{M}\right) \left\langle \hat{a}^{\dagger }%
\hat{b}^{\dagger }\right\rangle +i\Omega \left\langle \hat{b}^{\dagger
}\right\rangle -\frac{1}{2}\left( \Gamma _{a}+\Gamma _{b}\right)
\left\langle \hat{a}^{\dagger }\hat{b}^{\dagger }\right\rangle +ig_{opt}%
\left[ 2\left\langle \hat{a}^{\dagger }\hat{b}\right\rangle +\left\langle 
\hat{a}^{\dagger }\hat{b}^{\dagger }\hat{b}^{2}\right\rangle \right.  \notag
\\
&&\left. +2\left\langle \hat{a}^{\dagger 2}\hat{a}\hat{b}\right\rangle
+\left\langle \hat{a}^{\dagger }\hat{b}^{\dagger 3}\right\rangle +2\left(
\left\langle \hat{a}^{\dagger }\hat{b}^{\dagger }\right\rangle +\left\langle 
\hat{a}^{\dagger }\hat{b}^{\dagger 2}\hat{b}\right\rangle +\left\langle \hat{%
a}^{\dagger 2}\hat{a}\hat{b}^{\dagger }\right\rangle \right) +\left\langle 
\hat{a}^{\dagger }\hat{b}^{\dagger }\right\rangle \right].
\end{eqnarray}

\begin{equation}
\frac{d}{dt}\left\langle \hat{a}^{2}\right\rangle =-2i\Delta
_{c}\left\langle \hat{a}^{2}\right\rangle -2i\Omega \left\langle \hat{a}%
\right\rangle -\Gamma _{a}\left\langle \hat{a}^{2}\right\rangle -2ig_{opt}%
\left[ \left\langle \hat{a}^{2}\hat{b}^{\dagger 2}\right\rangle
+\left\langle \hat{a}^{2}\hat{b}^{2}\right\rangle +2\left\langle \hat{a}^{2}%
\hat{b}^{\dagger }\hat{b}\right\rangle +\left\langle \hat{a}%
^{2}\right\rangle \right].
\end{equation}

\begin{equation}
\frac{d}{dt}\left\langle \hat{a}^{\dagger 2}\right\rangle =2i\Delta
_{c}\left\langle \hat{a}^{\dagger 2}\right\rangle +2i\Omega \left\langle 
\hat{a}^{\dagger }\right\rangle -\Gamma _{a}\left\langle \hat{a}^{\dagger
2}\right\rangle +2ig_{opt}\left[ \left\langle \hat{a}^{\dagger 2}\hat{b}%
^{\dagger 2}\right\rangle +\left\langle \hat{a}^{\dagger 2}\hat{b}%
^{2}\right\rangle +2\left\langle \hat{a}^{\dagger 2}\hat{b}^{\dagger }\hat{b}%
\right\rangle +\left\langle \hat{a}^{\dagger 2}\right\rangle \right].
\end{equation}

\begin{equation}
\frac{d}{dt}\left\langle \hat{b}^{2}\right\rangle =-2i\omega
_{M}\left\langle \hat{b}^{2}\right\rangle -\Gamma _{b}\left\langle \hat{b}%
^{2}\right\rangle -2ig_{opt}\left[ \left\langle \hat{a}^{\dagger }\hat{a}%
\right\rangle +2\left\langle \hat{a}^{\dagger }\hat{a}\hat{b}^{\dagger }\hat{%
b}\right\rangle +2\left\langle \hat{a}^{\dagger }\hat{a}\hat{b}%
^{2}\right\rangle \right].
\end{equation}

\begin{equation}
\frac{d}{dt}\left\langle \hat{b}^{\dagger 2}\right\rangle =2i\omega
_{M}\left\langle \hat{b}^{\dagger 2}\right\rangle -\Gamma _{b}\left\langle 
\hat{b}^{\dagger 2}\right\rangle +2ig_{opt}\left[ \left\langle \hat{a}%
^{\dagger }\hat{a}\right\rangle +2\left\langle \hat{a}^{\dagger }\hat{a}\hat{%
b}^{\dagger }\hat{b}\right\rangle +2\left\langle \hat{a}^{\dagger }\hat{a}%
\hat{b}^{\dagger 2}\right\rangle \right].
\end{equation}
\newline
where $\overline{n}^{a}_{th}$, $\overline{n}^{b}_{th}$ are the thermal mean
number of cavity photons and membrane oscillations respectively.

\subsection{Coupled Equations and Decorrelation of Higher-Order Operators}

The above set of equations are not closed, constitute higher order operator
products, and need to be solved numerically with approximations. We proceed
to decorrelate all the higher-order correlations in the above equations. The
above set of equations will then be closed up to second order when we apply
the decorrelation method. We proceed to decorrelate the higher-(third-and
fourth-) order quantum correlations present in the above equations like
 \cite{Raymond}, which studied the correlation of photon
pairs from a double Raman Amplifier; a hybrid quantum optomechanical system
containing a single semiconductor quantum well \cite{Singh}. This approach
corresponds to truncation of higher-order operator products in order to
solve for all the second-order correlation functions. A similar kind of
approximation has also been used in Ref. \cite{Anglin} to study the dynamics
of a two-mode BEC beyond mean field approximation:

\begin{equation}
\left\langle ABC\right\rangle \approx \lbrack \left\langle A\right\rangle
\left\langle BC\right\rangle +\left\langle AB\right\rangle \left\langle
C\right\rangle +\left\langle AC\right\rangle \left\langle B\right\rangle ]
\end{equation}
and 
\begin{equation}
\left\langle ABCD\right\rangle \approx \lbrack \left\langle AB\right\rangle
\left\langle CD\right\rangle +\left\langle AC\right\rangle \left\langle
BD\right\rangle +\left\langle AD\right\rangle \left\langle BC\right\rangle ].
\end{equation}

\subsection{Temporal Evolutions}

After applying the decorrelation method, the above set of equations reduced
to a closed set of coupled equations given in Appendix A. These equations
contain all possible first-order operator averages and second-order
correlation functions. The closed set of equations, given in Appendix A, is
further solved numerically for the transient dynamics as well as
nonclassical photon statistics.\ 

Figure 1 shows the temporal evolution of the mean number of excitations for
the cavity photons $\left\langle\hat{a}^{\dagger}\hat{a}\right\rangle$ and
phonons due to mechanical motion of moving membrane $\left\langle\hat{b}%
^{\dagger}\hat{b}\right\rangle$ for a set of different detuning parameters.
For a detuning of the order of mechanical frequency $\omega _{M}$, cavity is
driven around red sideband associated with the membrane oscillator and we
have photonic oscillations $\left\langle\hat{a}^{\dagger}\hat{a}\right\rangle
$ with phonon sidebands $\left\langle \hat{b}^{\dagger }\hat{b}\right\rangle 
$. Both mean photonic and phonon excitations are asymmetric in shape also
due to strong quadratic optomechanical coupling. As the detuning increases,
cavity is driven far from red sideband regime due to which amplitude of the
oscillations for $\left\langle\hat{a}^{\dagger }\hat{a}\right\rangle$ are
reduced as shown in Figure 1(c). For a very large detuned cavity,
oscillations for $\left\langle \hat{a}^{\dagger }\hat{a}\right\rangle $ is
completely reduced to its initial value, whereas for $\left\langle \hat{b}%
^{\dagger }\hat{b}\right\rangle $ we have a regular oscillations over the
whole range of time scale. So, for a quadratic coupled optomechanical system
cavity should be driven in red (blue) sideband regime of membrane
oscillator. This is also required for generation of sub-Poissonian light
from the driven cavity in our system Hamiltonian.\ 

In Figure 2, we observe the effect of cavity decay on the temporal dynamics
of the system. In the weak cavity decay regime, we have undamped
oscillations for photonic as well as phonon oscillations as shown in Figure
2(a). However, in the strong cavity decay regime, amplitude of oscillations
for both $\left\langle \hat{a}^{\dagger}\hat{a}\right\rangle $ as well as $%
\left\langle \hat{b}^{\dagger}\hat{b}\right\rangle$ show  damping with time
as shown in the Figure 2(b). Although, in both the cases decay rate for
phonons $\Gamma _{b}$ remains the same. For strong cavity decay, photons
leak out of the cavity very rapidly and this leads to damped oscillations
for mean phonon number $\left\langle \hat{b}^{\dagger}\hat{b}\right\rangle$
also with time.\ 

We also study the temporal dynamics for finite thermal phonon numbers in
Figure 3. We can see that the mean phonon number $\left\langle \hat{b}%
^{\dagger}\hat{b}\right\rangle$ due to mechanical motion of membrane
increases and saturates with time due to thermal noise (for finite $%
\overline{n}_{th}^{b}$) even for the strong decay regime of cavity as well
as mechanical mode as shown in Figure 3(b). So, a finite thermal noise due
to phonons is found to increase the mean number of excitations for the
dielectric membrane. This is due to dependence of mean excitations number of
phonons $\left(\left\langle \hat{b}^{\dagger}\hat{b}\right\rangle\right)$ on
thermal noise and can be also seen in Eq.{A6}. In this case mean excitations
number for photonic and phonons also oscillate in opposite phase upto
certain range of time. 
\begin{figure}[hbt!]
\centering
\subfloat[]{\includegraphics[width=0.5\textwidth]{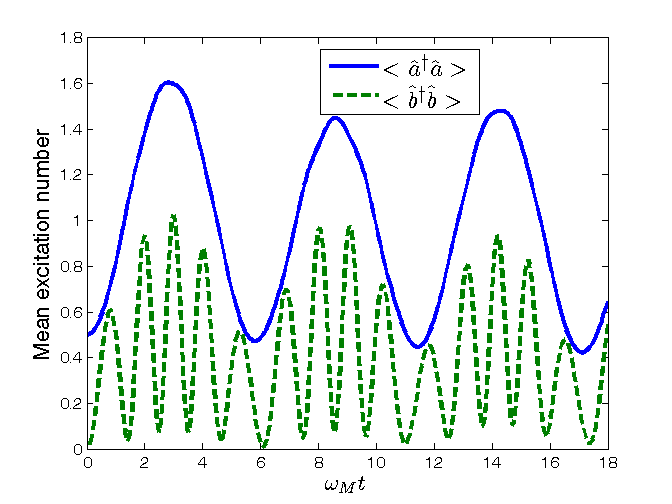}} \subfloat[]{%
\includegraphics[width=0.5\textwidth]{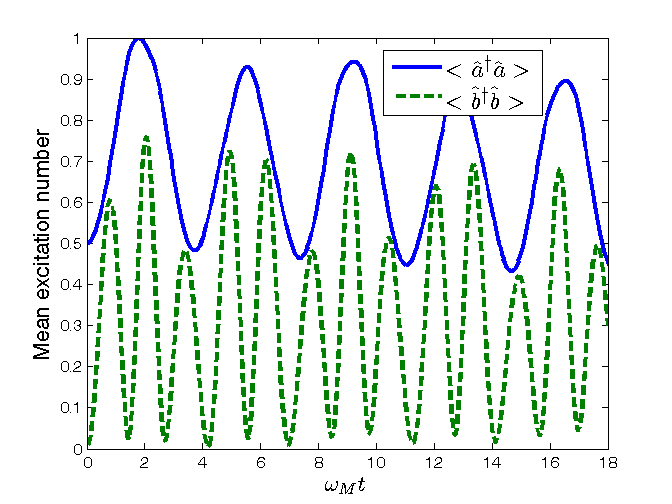}} \newline
\subfloat[]{\includegraphics[width=0.5\textwidth]{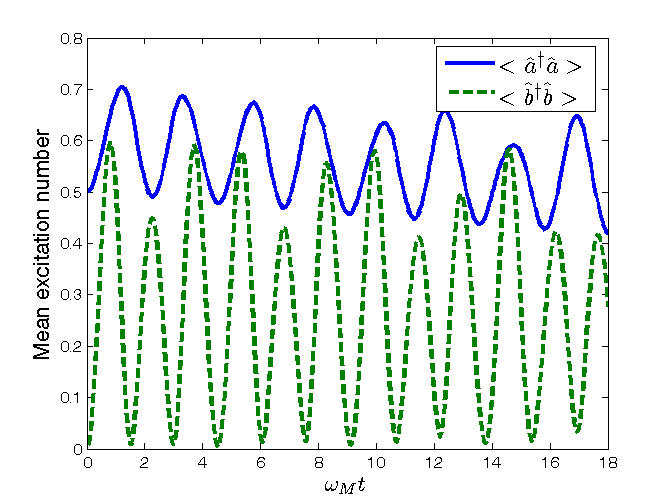}} \subfloat[]{%
\includegraphics[width=0.5\textwidth]{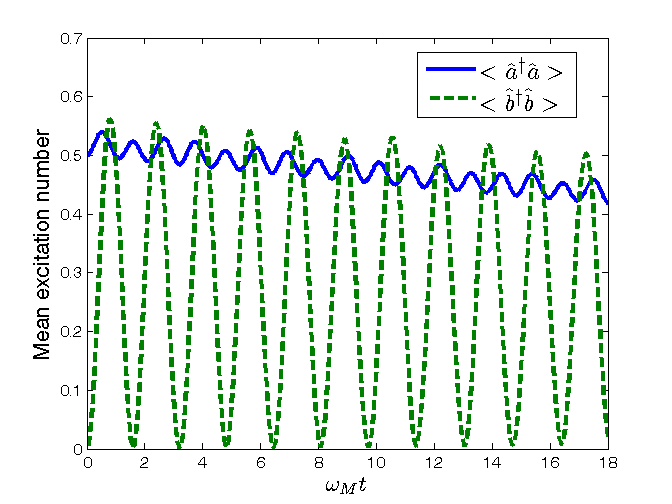}} \newline
\caption{(Color online) Mean number of cavity photons $\left\langle \hat{a}
^{\dagger}\hat{a}\right\rangle$ (blue line) and phonons of moving membrane $%
\left\langle \hat{b}^{\dagger}\hat{b}\right\rangle$ (green line) for chosen
set of parameters $g_{opt}/\protect\omega_{M}= 1.4$; $\Gamma _{a}/\protect%
\omega _{M}=0.01$; $\Gamma_{b}/\protect\omega _{M}=0.001$; $\Omega /\protect%
\omega _{M}=0.6$; $\overline{n}^{a}_{th} = \overline{n}^{b}_{th}= 0$; where $%
\protect\omega _{M}$ is frequency of mechanical motion of the membrane
chosen to 1 in our numerical simulations. a) $\Delta _{c}/\protect\omega %
_{M}= 0.5$; b) $\Delta _{c}/\protect\omega _{M}= 1.0$; c) $\Delta _{c}/%
\protect\omega _{M}= 2.0$; d) $\Delta _{c}/\protect\omega _{M}= 5.0$.}
\label{Fig 1}
\end{figure}
\begin{figure}[hbt!]
\centering
\subfloat[]{\includegraphics[width=0.5\textwidth]{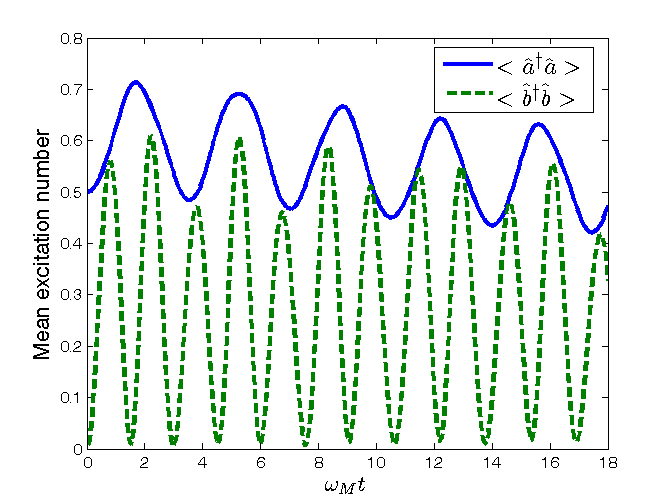}} \subfloat[]{%
\includegraphics[width=0.5\textwidth]{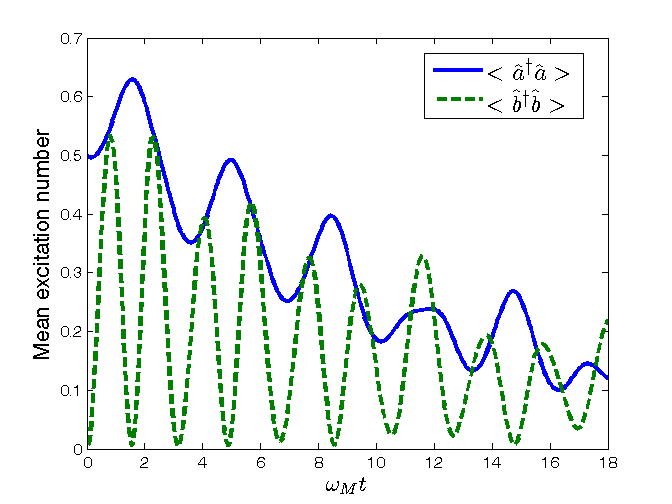}} \newline
\caption{(Color online) Mean number of cavity photons $\left\langle \hat{a}
^{\dagger}\hat{a}\right\rangle$ (blue line) and phonons of moving membrane $%
\left\langle \hat{b}^{\dagger}\hat{b}\right\rangle$ (green line) for chosen
set of parameters $\Delta _{c}/\protect\omega _{M}= 1.0$; $g_{opt}/\protect%
\omega_{M}= 1.4$; $\Gamma_{b}/\protect\omega _{M}=0.001$; $\Omega /\protect%
\omega _{M}=0.4$; $\overline{n}^{a}_{th} = \overline{n}^{b}_{th}= 0$; a)For
weak cavity decay $\Gamma _{a}/\protect\omega _{M}=0.01$; b) For strong
cavity decay $\Gamma _{a}/\protect\omega _{M}=0.1$.}
\label{Fig 2}
\end{figure}
\begin{figure}[hbt!]
\centering
\subfloat[]{\includegraphics[width=0.5\textwidth]{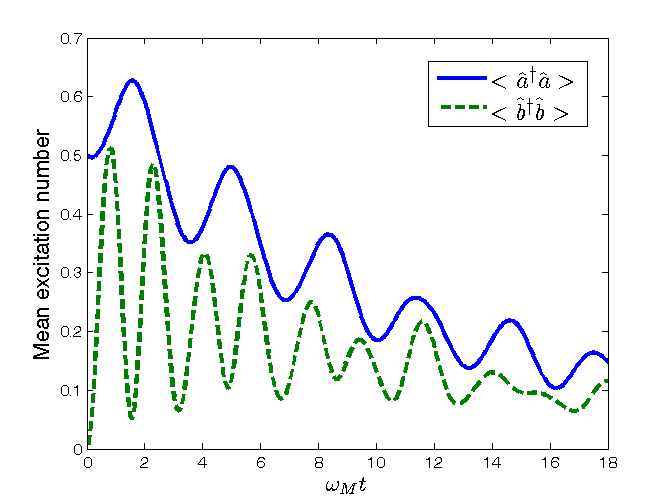}} \subfloat[]{%
\includegraphics[width=0.5\textwidth]{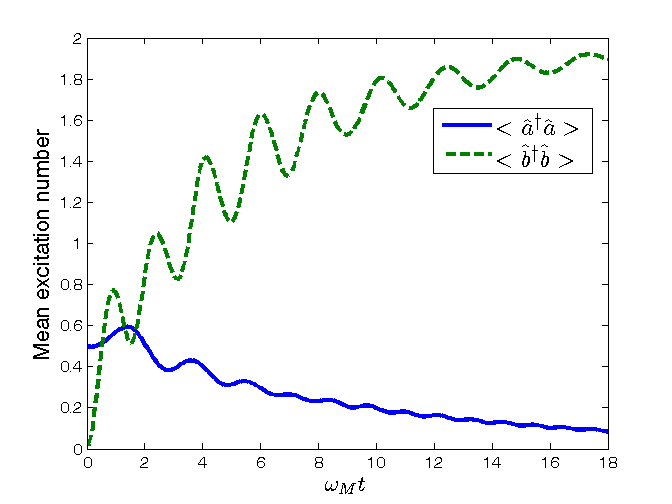}} \newline
\caption{(Color online) Effects of thermal phonons on the mean number of
cavity photons $\left\langle \hat{a}^{\dagger}\hat{a}\right\rangle$ (blue
line) and phonons of moving membrane $\left\langle \hat{b}^{\dagger}\hat{b}%
\right\rangle$ (green line) for chosen set of parameters $\Delta _{c}/%
\protect\omega _{M}= 1.0$; $g_{opt}/\protect\omega_{M}= 1.4$; $\Omega /%
\protect\omega _{M}=0.4$ under strong decay regime $\Gamma _{a}/\protect%
\omega _{M} = \Gamma_{b}/\protect\omega _{M}=0.1$; a) For $\overline{n}%
^{b}_{th}= 0.0$; b) $\overline{n}^{b}_{th}= 2.0$.}
\label{Fig 3}
\end{figure}

\section{Photon Statistics}

In addition to the mean excitation numbers for cavity $\left(\left\langle%
\hat{a}^{\dagger}\hat{a}\right\rangle \right)$ as well as moving membrane $%
\left( \left\langle \hat{b}^{\dagger}\hat{b}\right\rangle \right)$, we have
computed the normalized two-boson correlation functions for the cavity field 
$g_{a}^{\left(2\right)}(0)$, quadratically coupled membrane $%
g_{b}^{\left(2\right)}(0)$ , and the cross-correlation between them $%
g_{ab}^{\left(2\right)}(0)$ using \cite{Scully,Raymond,Singh}. 
\begin{equation}
g_{a}^{\left(2\right)}(0)=\frac{\left\langle \hat{a}^{\dagger 2}\hat{a}%
^{2}\right\rangle }{\left\langle \hat{a}^{\dagger }\hat{a}\right\rangle ^{2}}%
\approx \left[ \frac{2\left\langle \hat{a}^{\dagger }\hat{a}\right\rangle
\left\langle \hat{a}^{\dagger }\hat{a}\right\rangle +\left\langle \hat{a}%
^{\dagger 2}\right\rangle \left\langle \hat{a}^{2}\right\rangle }{%
\left\langle \hat{a}^{\dagger }\hat{a}\right\rangle ^{2}}\right] .
\end{equation}%
\begin{equation}
g_{b}^{\left(2\right)}(0)=\frac{\left\langle \hat{b}^{\dagger 2}\hat{b}%
^{2}\right\rangle }{\left\langle \hat{b}^{\dagger }\hat{b}\right\rangle ^{2}}%
\approx \left[ \frac{2\left\langle \hat{b}^{\dagger }\hat{b}\right\rangle
\left\langle \hat{b}^{\dagger }\hat{b}\right\rangle +\left\langle \hat{b}%
^{\dagger 2}\right\rangle \left\langle \hat{b}^{2}\right\rangle }{%
\left\langle \hat{b}^{\dagger }\hat{b}\right\rangle ^{2}}\right] .
\end{equation}%
\begin{equation}
g_{ab}^{\left( 2\right)}(0)=\frac{\left\langle \hat{a}^{\dagger }\hat{b}%
^{\dagger }\hat{b}\hat{a}\right\rangle }{\left\langle \hat{b}^{\dagger }\hat{%
b}\right\rangle \left\langle \hat{a}^{\dagger }\hat{a}\right\rangle }\approx %
\left[ \frac{\left\langle \hat{a}^{\dagger }\hat{b}\right\rangle
\left\langle \hat{b}^{\dagger }\hat{a}\right\rangle +\left\langle \hat{b}%
^{\dagger }\hat{b}\right\rangle \left\langle \hat{a}^{\dagger }\hat{a}%
\right\rangle +\left\langle \hat{a}^{\dagger }\hat{b}^{\dagger
}\right\rangle \left\langle \hat{b}\hat{a}\right\rangle }{\left\langle \hat{b%
}^{\dagger }\hat{b}\right\rangle \left\langle \hat{a}^{\dagger }\hat{a}%
\right\rangle }\right] .
\end{equation}
\newline
If $g_{a}^{\left( 2\right) }(0)$, $g_{b}^{\left( 2\right) )}(0)$ and $%
g_{ab}^{\left( 2\right) }(0)$ satisfy the inequality \ $g_{X}^{\left(
2\right) }(0)<1$; $X=a,b,ab$, then the statistics of the bosonic systems are
referred to as \textit{sub-Poissonian}. Statistics with $\
g_{X}^{\left(2\right)}(0)=1$ and $g_{X}^{\left( 2\right)}(0)>1$ are
similarly referred to as\textit{\ Poissonian} and \textit{super-Poissonian,}
respectively. Here, we have studied the temporal dynamics of second-order
correlation functions, namely the self-correlation and cross correlation for
cavity and oscillating membrane modes. \newline

In photon blockade effect, where coupling of a single photon to the system
hinders the coupling of the subsequent photons, we have $g_{a}^{\left(
2\right)}(0)<1$ for the cavity mode. Similarly, in photon induced tunneling
regime, coupling of initial photons favors the coupling of the subsequent
photons and leads to condition $g_{a}^{\left( 2\right) }(0)>1$. \ We examine
the effects of various parameters on the temporal behavior of $g_{a}^{\left(
2\right) }(0)$, $g_{b}^{\left( 2\right) }(0)$ and $g_{ab}^{\left( 2\right)
}(0)$ as well as under different decay regimes for cavity photons as well as
phonons due to micromechanical motion.\newline

The second order correlation functions, namely the self- and
cross-correlations for photonic and membrane oscillations are shown in
Figure 4. We study the temporal dynamics of $g_{a}^{\left(2\right)}(0)$, $%
g_{b}^{\left(2\right))}(0)$ and $g_{ab}^{\left(2\right)}(0)$ for different
cavity detunings $\Delta _{c}$. It can be seen that for a resonant cavity
i.e. $\Delta _{c}=0$ and for a shorter period of time $\omega_{M}t\sim 2$,
all the three correlations follow sub-Poissonian photon statistics as shown
in Figure 4 (a). As $\omega _{M}t$ increases, $g_{a}^{\left(2\right)}(0)$
tends to become super-Poissonian, whereas $g_{b}^{\left(2\right)}(0)$
oscillates from sub-Poissonian to super-Poissonian. For a finite detuning $%
\Delta _{c}\sim \omega_{M}$, $g_{a}^{\left(2\right)}(0)$ as well as $%
g_{b}^{\left(2\right))}(0)$ both oscillate periodically from sub-Poissonian
to super-Poissonian whereas $g_{ab}^{\left(2\right)}(0)$ mostly remains
sub-Poissonian as shown in Figure 4(b). For a very far off detuned cavity, $%
g_{a}^{\left(2\right)}(0)$ oscillates from Poissonian to super-Poissonian
and it never becomes sub-Poissonian, whereas $g_{b}^{\left(2\right))}(0)$
behaves almost the same and $g_{ab}^{\left(2\right))}(0)$ oscillates from
sub-Poissonian to Poissonian as shown in Figure 4(c) and Figure 4(d). For a
very large cavity detuning, a single photon can not be resonantly excited
into the cavity, while the probability of finding two or more photons
resonantly enhanced \cite{Nori}. So, we do not have photon blockade effect
for larger cavity detuning at any period of time $\omega_{M}t$.\ 

Similarly, for a finite detuning $\Delta _{c}$, we study the effect of
optomechanical coupling strengths $g_{opt}$ on various correlations in
Figure 5. For the case $g_{opt}$ comparable to driving field $\ \Omega $, $\
g_{a}^{\left( 2\right) )}(0)$ and $g_{b}^{\left( 2\right) )}(0)$ oscillate
from sub-Poissonian to super-Poissonian, whereas $g_{ab}^{\left( 2\right)
)}(0)$ remains mostly sub-Poissonian except at some points as shown in
Figure 5(a) and Figure 5(b). For a very strong optomechanical coupling $\
g_{opt}$, $\ g_{a}^{\left( 2\right) )}(0)$ oscillates from Poissonian to
super-Poissonian, whereas $g_{b}^{\left( 2\right) )}(0)$ remains same and $%
g_{ab}^{\left( 2\right) )}(0)$ varies to Poissonian limit as shown in Figure
5(d). \ 

Furthermore, we examine the effects of photonic and phonon decay
rates on various two-boson correlations in Figure 6. For very weak decay
rates of cavity and membrane oscillation, $g_{a}^{\left( 2\right) )}(0)$
varies from Poissonian to super-Poissonian and never becomes sub-Poissonian.
Although $g_{b}^{\left( 2\right) )}(0)$ oscillates from sub-Poissonian to
super-Poissonian and $g_{ab}^{\left( 2\right) )}(0)$ remains mostly
sub-Poissonian as shown in Figure 6(a). For a strong cavity decay rate $%
\Gamma _{a}$, $g_{a}^{\left( 2\right) )}(0)$ becomes sub-Poissonian over a
large scale of time and even becomes smaller than $0.5$ as shown in Figure
6(b). This means, we have a photon blockade effect only in resolved sideband
regime $\Gamma _{a}/\omega _{M}< 1$ but not in the deep-resolved-sideband
regime $\Gamma _{a}/\omega _{M}<< 1$ . While, for a very strong cavity as
well as membrane oscillation decay rates, all the three correlations  decay with time
rapidly as shown in Figure 6(c). So, strong photonic and
phonon decay rates play a positive role for getting sub-Poissonian photon
statistics from a driven cavity in our system Hamiltonian. 
\begin{figure}[hbt!]
\centering
\subfloat[]{\includegraphics[width=0.5\textwidth]{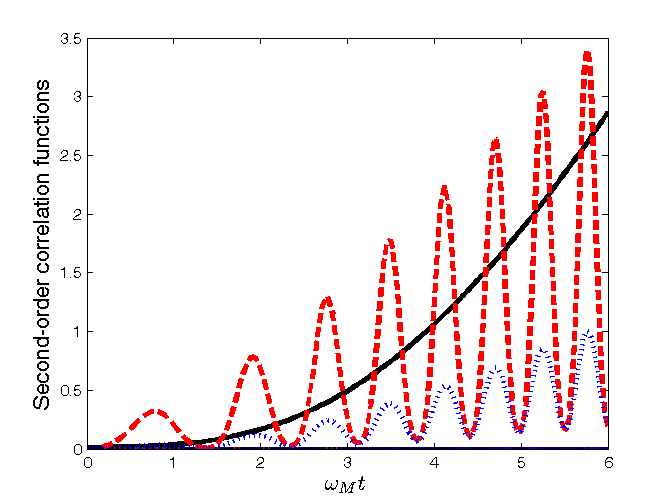}} \subfloat[]{%
\includegraphics[width=0.5\textwidth]{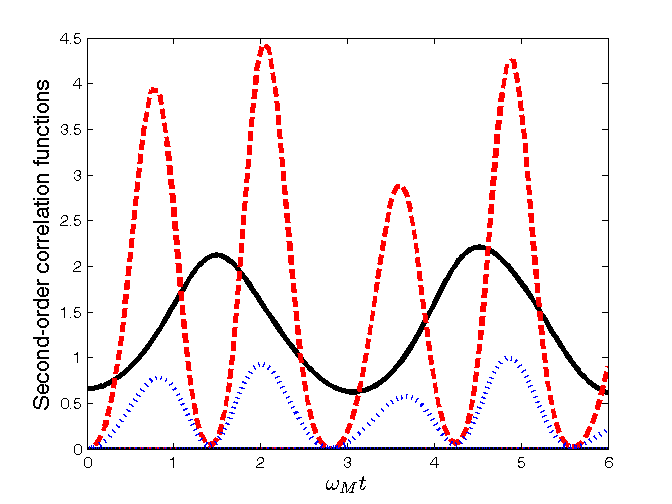}} \newline
\subfloat[]{\includegraphics[width=0.5\textwidth]{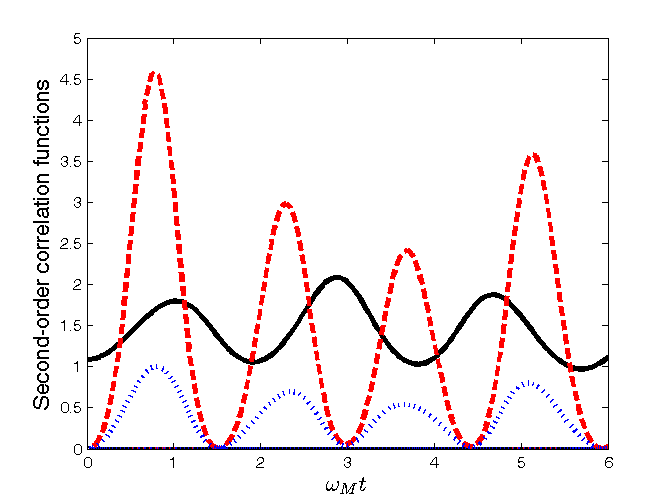}} \subfloat[]{%
\includegraphics[width=0.5\textwidth]{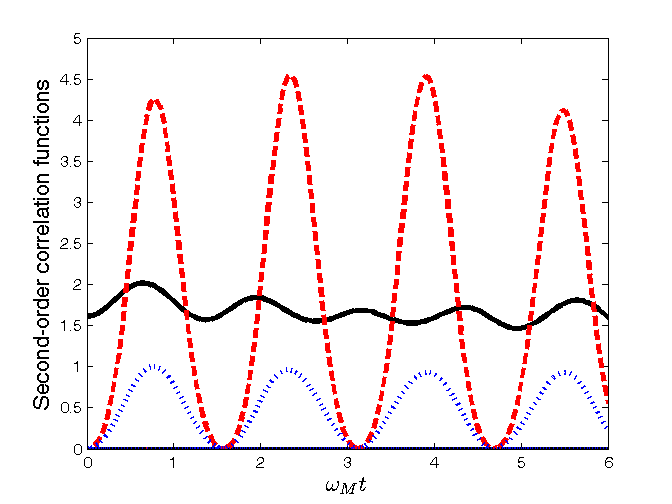}} \newline
\caption{Second-order autocorrelations $g_{a}^{\left( 2\right) }\left(
0\right) $ (black solid line) for cavity mode, $g_{b}^{\left( 2\right)
}\left( 0\right) $ for moving membrane (red dashed line) and $g_{ab}^{\left(
2\right) }\left( 0\right) $ (blue dotted line) corresponding
cross-correlation between photons and phonons for chosen set of parameters $%
g_{opt}/\protect\omega _{M}=1.5$; $\Gamma _{a}/\protect\omega _{M}=0.01$; $%
\Gamma _{b}/\protect\omega _{M}=0.001$; $\Omega /\protect\omega _{M}=0.6$; $%
\overline{n}_{th}^{a}=\overline{n}_{th}^{b}=0$; a) $\Delta _{c}/\protect%
\omega _{M}=0.0$; b) $\Delta _{c}/\protect\omega _{M}=1.3$; c) $\Delta _{c}/%
\protect\omega _{M}=2.5$; d) $\Delta _{c}/\protect\omega _{M}=4.0$.}
\label{Fig 4}
\end{figure}
\begin{figure}[hbt!]
\centering
\subfloat[]{\includegraphics[width=0.5\textwidth]{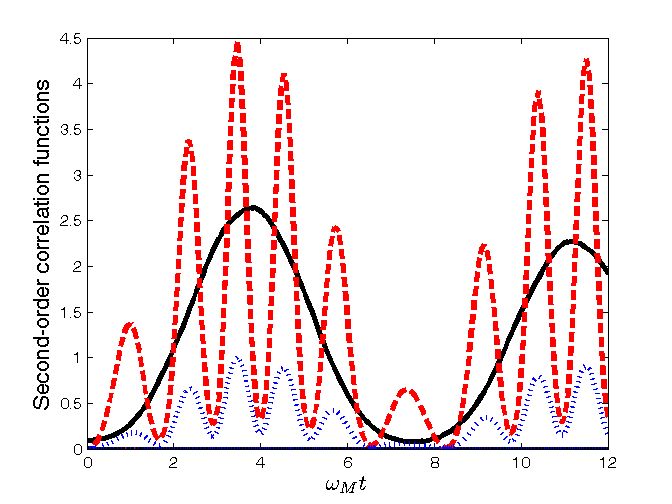}} \subfloat[]{%
\includegraphics[width=0.5\textwidth]{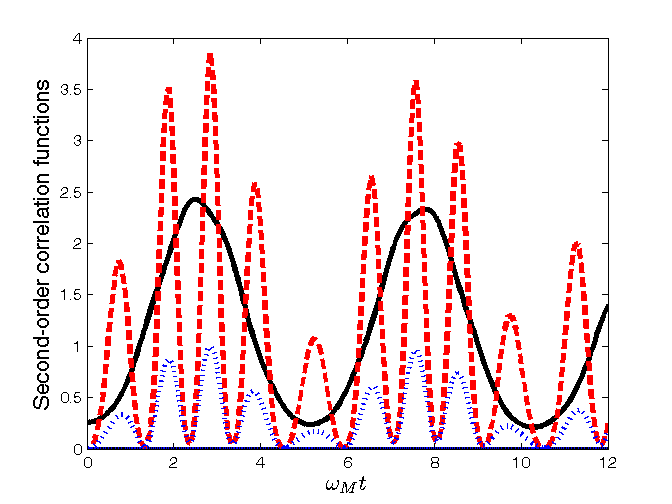}} \newline
\subfloat[]{\includegraphics[width=0.5\textwidth]{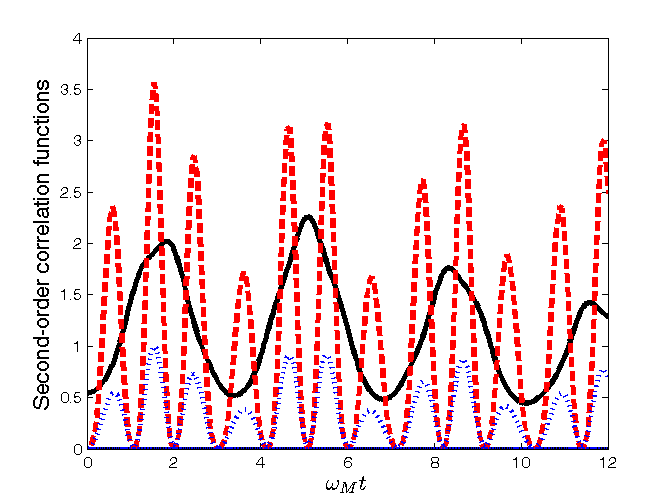}} \subfloat[]{%
\includegraphics[width=0.5\textwidth]{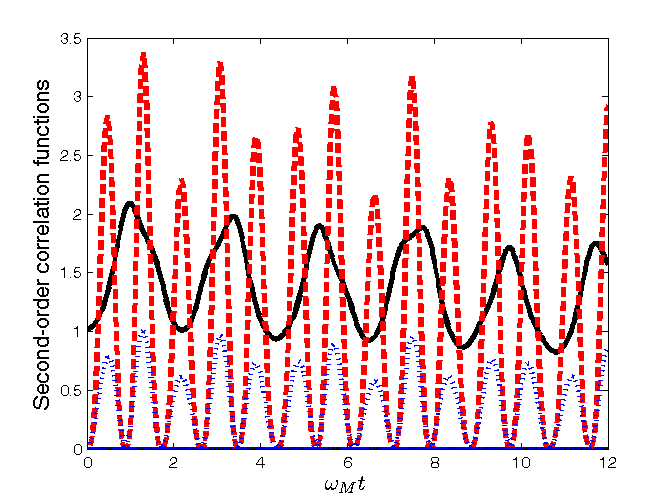}} \newline
\caption{Second-order autocorrelations $g_{a}^{\left( 2\right) }\left(
0\right) $ (black solid line) for cavity mode, $g_{b}^{\left( 2\right)
}\left( 0\right) $ for moving membrane (red dashed line) and $g_{ab}^{\left(
2\right) }\left( 0\right) $ (blue dotted line) corresponding
cross-correlation between photons and phonons for chosen set of parameters $%
\Delta _{c}/\protect\omega _{M}=0.5$; $\Gamma _{a}/\protect\omega _{M}=0.01$%
; $\Gamma _{b}/\protect\omega _{M}=0.001$; $\Omega /\protect\omega _{M}=0.6$%
; $\overline{n}_{th}^{a}=\overline{n}_{th}^{b}=0$; a) $g_{opt}/\protect%
\omega _{M}=0.8$; b) $g_{opt}/\protect\omega _{M}=1.7$; c) $g_{opt}/\protect%
\omega _{M}=3.0$; d) $g_{opt}/\protect\omega _{M}=5.0$.}
\label{Fig 5}
\end{figure}
\begin{figure}[hbt!]
\centering
\subfloat[]{\includegraphics[width=0.5\textwidth]{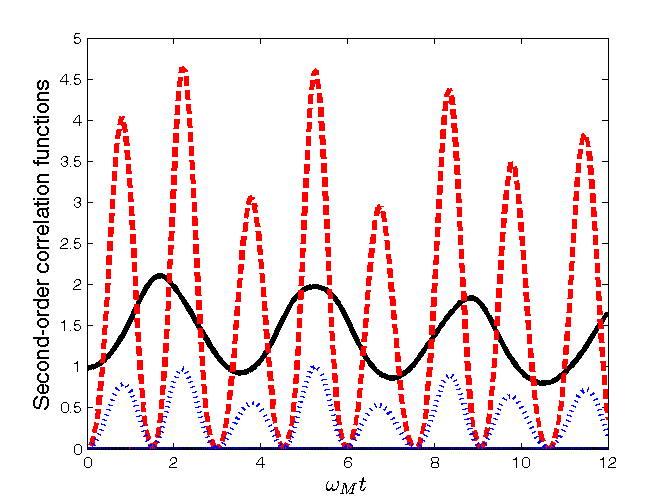}} \subfloat[]{%
\includegraphics[width=0.5\textwidth]{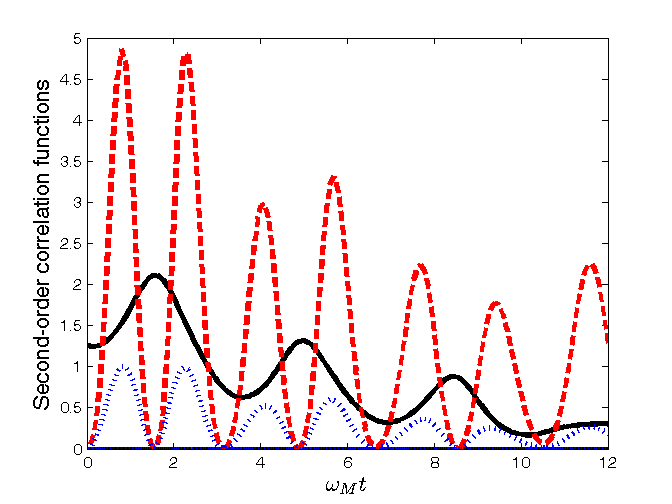}} \newline
\subfloat[]{\includegraphics[width=0.5\textwidth]{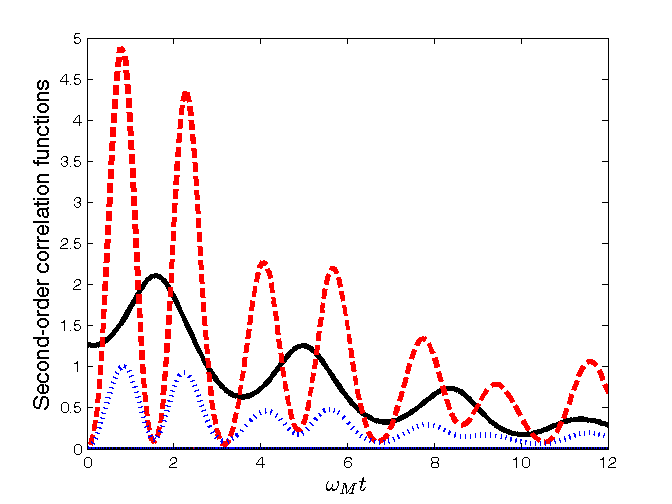}}
\caption{Second-order autocorrelations $g_{a}^{\left( 2\right) }\left(
0\right) $ (black solid line) for cavity mode, $g_{b}^{\left( 2\right)
}\left( 0\right) $ for moving membrane (red dashed line) and $g_{ab}^{\left(
2\right) }\left( 0\right) $ (blue dotted line) corresponding
cross-correlation between photons and phonons for chosen set of parameters $%
\Delta _{c}/\protect\omega _{M}=1.0$; $g_{opt}/\protect\omega _{M}=1.4$; $%
\Omega /\protect\omega _{M}=0.4$; $\overline{n}_{th}^{a}=\overline{n}%
_{th}^{b}=0$ under different decay regimes; a) $\Gamma _{a}/\protect\omega %
_{M}=0.01$, $\Gamma _{b}/\protect\omega _{M}=0.001$; b) $\Gamma _{a}/\protect%
\omega _{M}=0.1$; $\Gamma _{b}/\protect\omega _{M}=0.001$; c) $\Gamma _{a}/%
\protect\omega _{M}=\Gamma _{b}/\protect\omega _{M}=0.1$.}
\label{Fig 6}
\end{figure}

\section{Conclusion}

We have studied a qudratically coupled optomechanical system formed by a micropillar with
moveable Bragg reflectors and a thin dielectric membrane at the node (or
antinode) of the resonator through the Heisenberg- Langevin approach. We
have obtained coupled equations up to the second order without any
approximation. We decorrelate the higher-order correlation
functions in the coupled equations  to obtain a closed set of equations.
This enables  the study of  temporal dynamics of the optomechanical system
as well as two-boson correlation functions under different scenarios. In the
resolved sideband regime, mean number of cavity as well as mechanical
excitation show oscillations. Furthermore, amplitude of these oscillations changes as we
change the cavity detuning. We have also studied the temporal dynamics under
the different decay regimes of cavity and membrane oscillations. We have
discussed the effect of thermal noise due to phonons on the dynamics of
the optomechanical system. \ Furthermore, we analyzed the temporal dynamics
of the second-order correlation functions, namely the self and
cross-correlation of the cavity and mechanical modes under different cavity
detunings as well as optomechanical coupling strengths. We obtained the
regimes where all the three correlations display strong sub-Poissonian
photon statistics particularly at very small cavity detuning as well as
strong cavity decay rate. Our study is useful
for coherent control of photon statistics as well as photon and phonon
correlations in quadratically coupled optomechanical systems.

\section*{Acknowledgments}

The authors thank University of Malaya for financial support under the
University of Malaya Research Grant (RG231-12AFR and RP006A-13AFR). This work also get supported through the postdoctoral fellowship position held by Shailendra Kumar Singh in T\"{U}B\.{I}TAK-1001  Grant No.  114F170. SKS would like to gratefully acknowledge Dr. Mehmet  Emre Tasgin  for his kind help to introduce the subject of cavity optomechanics in details. SKS also kindly acknowledge to Hacettepe university colleague Mr. Saidul Alom Mozumdar for his help to improve manuscript grammatically.

\appendix

\section{Decorrelated and Closed set of Coupled Equations}

\begin{eqnarray}
\frac{d}{dt}\left\langle \hat{a}\right\rangle &=&-i\Delta _{c}\left\langle 
\hat{a}\right\rangle -i\Omega -\frac{1}{2}\Gamma _{a}\left\langle \hat{a}%
\right\rangle -ig_{opt}\left[ \left\{ \left\langle \hat{a}\right\rangle
\left\langle \hat{b}^{\dagger 2}\right\rangle +2\left\langle \hat{a}\hat{b}%
^{\dagger }\right\rangle \left\langle \hat{b}^{\dagger }\right\rangle
\right\} \right. \\
&&\left. +\left\{ \left\langle \hat{a}\right\rangle \left\langle \hat{b}%
^{2}\right\rangle +2\left\langle \hat{a}\hat{b}\right\rangle \left\langle 
\hat{b}\right\rangle \right\} +2\left\{ \left\langle \hat{a}\right\rangle
\left\langle \hat{b}^{\dagger }\hat{b}\right\rangle +\left\langle \hat{a}%
\hat{b}^{\dagger }\right\rangle \left\langle \hat{b}\right\rangle
+\left\langle \hat{a}\hat{b}\right\rangle \left\langle \hat{b}^{\dagger
}\right\rangle \right\} +\left\langle \hat{a}\right\rangle \right] .  \notag
\end{eqnarray}%
\begin{eqnarray}
\frac{d}{dt}\left\langle \hat{a}^{\dagger }\right\rangle &=&i\Delta
_{c}\left\langle \hat{a}^{\dagger }\right\rangle +i\Omega -\frac{1}{2}\Gamma
_{a}\left\langle \hat{a}^{\dagger }\right\rangle +ig_{opt}\left[ \left\{
\left\langle \hat{a}^{\dagger }\right\rangle \left\langle \hat{b}^{\dagger
2}\right\rangle +2\left\langle \hat{a}^{\dagger }\hat{b}^{\dagger
}\right\rangle \left\langle \hat{b}^{\dagger }\right\rangle \right\} \right.
\\
&&\left. +\left\{ \left\langle \hat{a}^{\dagger }\right\rangle \left\langle 
\hat{b}^{2}\right\rangle +2\left\langle \hat{a}^{\dagger }\hat{b}%
\right\rangle \left\langle \hat{b}\right\rangle \right\} +2\left\{
\left\langle \hat{a}^{\dagger }\right\rangle \left\langle \hat{b}^{\dagger }%
\hat{b}\right\rangle +\left\langle \hat{a}^{\dagger }\hat{b}^{\dagger
}\right\rangle \left\langle \hat{b}\right\rangle +\left\langle \hat{a}%
^{\dagger }\hat{b}\right\rangle \left\langle \hat{b}^{\dagger }\right\rangle
\right\} +\left\langle \hat{a}^{\dagger }\right\rangle \right] .  \notag
\end{eqnarray}%
\begin{eqnarray}
\frac{d}{dt}\left\langle \hat{b}\right\rangle &=&-i\omega _{M}\left\langle 
\hat{b}\right\rangle -\frac{1}{2}\Gamma _{b}\left\langle \hat{b}%
\right\rangle -2ig_{opt}\left[ \left\langle \hat{a}^{\dagger }\right\rangle
\left\{ \left\langle \hat{a}\hat{b}\right\rangle +\left\langle \hat{a}\hat{b}%
^{\dagger }\right\rangle \right\} \right.  \notag \\
&&\left. +\left\langle \hat{a}^{\dagger }\hat{a}\right\rangle \left\{
\left\langle \hat{b}\right\rangle +\left\langle \hat{b}^{\dagger
}\right\rangle \right\} +\left\langle \hat{a}\right\rangle \left\{
\left\langle \hat{a}^{\dagger }\hat{b}\right\rangle +\left\langle \hat{a}%
^{\dagger }\hat{b}^{\dagger }\right\rangle \right\} \right] .
\end{eqnarray}%
\begin{eqnarray}
\frac{d}{dt}\left\langle \hat{b}^{\dagger }\right\rangle &=&i\omega
_{M}\left\langle \hat{b}^{\dagger }\right\rangle -\frac{1}{2}\Gamma
_{b}\left\langle \hat{b}^{\dagger }\right\rangle +2ig_{opt}\left[
\left\langle \hat{a}^{\dagger }\right\rangle \left\{ \left\langle \hat{a}%
\hat{b}\right\rangle +\left\langle \hat{a}\hat{b}^{\dagger }\right\rangle
\right\} \right.  \notag \\
&&\left. +\left\langle \hat{a}^{\dagger }\hat{a}\right\rangle \left\{
\left\langle \hat{b}\right\rangle +\left\langle \hat{b}^{\dagger
}\right\rangle \right\} +\left\langle \hat{a}\right\rangle \left\{
\left\langle \hat{a}^{\dagger }\hat{b}\right\rangle +\left\langle \hat{a}%
^{\dagger }\hat{b}^{\dagger }\right\rangle \right\} \right] .
\end{eqnarray}%
\begin{equation}
\frac{d}{dt}\left\langle \hat{a}^{\dagger }\hat{a}\right\rangle =-i\Omega
\left( \left\langle \hat{a}^{\dagger }\right\rangle -\left\langle \hat{a}%
\right\rangle \right) -\Gamma _{a}\left\langle \hat{a}^{\dagger }\hat{a}%
\right\rangle +\Gamma _{a}\bar{n}_{th}^{a}.
\end{equation}%
\begin{eqnarray}
\frac{d}{dt}\left\langle \hat{b}^{\dagger }\hat{b}\right\rangle &=&-2ig_{opt}%
\left[ \left\{ \left\langle \hat{a}^{\dagger }\hat{a}\right\rangle
\left\langle \hat{b}^{\dagger 2}\right\rangle +2\left\langle \hat{a}%
^{\dagger }\hat{b}^{\dagger }\right\rangle \left\langle \hat{a}\hat{b}%
^{\dagger }\right\rangle \right\} -\left\{ \left\langle \hat{a}^{\dagger }%
\hat{a}\right\rangle \left\langle \hat{b}^{2}\right\rangle +2\left\langle 
\hat{a}^{\dagger }\hat{b}\right\rangle \left\langle \hat{a}\hat{b}%
\right\rangle \right\} \right]  \notag \\
&&-\Gamma _{b}\left\langle \hat{b}^{\dagger }\hat{b}\right\rangle +\Gamma
_{b}\bar{n}_{th}^{b}.
\end{eqnarray}%
\begin{eqnarray}
\frac{d}{dt}\left\langle \hat{a}\hat{b}^{\dagger }\right\rangle &=&i\left(
\omega _{M}-\Delta _{c}\right) \left\langle \hat{a}\hat{b}^{\dagger
}\right\rangle -i\Omega \left\langle \hat{b}^{\dagger }\right\rangle -\frac{1%
}{2}\left( \Gamma _{a}+\Gamma _{b}\right) \left\langle \hat{a}\hat{b}%
^{\dagger }\right\rangle  \notag \\
&&+ig_{opt}\left[ 2\left\{ 2\left\langle \hat{a}^{\dagger }\hat{a}%
\right\rangle \left\langle \hat{a}\hat{b}\right\rangle +\left\langle \hat{a}%
^{\dagger }\hat{b}\right\rangle \left\langle \hat{a}^{2}\right\rangle
\right\} -\left\{ \left\langle \hat{a}\hat{b}^{\dagger }\right\rangle
\left\langle \hat{b}^{2}\right\rangle +2\left\langle \hat{a}\hat{b}%
\right\rangle \left\langle \hat{b}^{\dagger }\hat{b}\right\rangle \right\}
\right.  \notag \\
&&+2\left\{ \left( 2\left\langle \hat{a}^{\dagger }\hat{a}\right\rangle
\left\langle \hat{a}\hat{b}^{\dagger }\right\rangle +\left\langle \hat{a}%
^{\dagger }\hat{b}^{\dagger }\right\rangle \left\langle \hat{a}%
^{2}\right\rangle \right) -\left( 2\left\langle \hat{b}^{\dagger }\hat{b}%
\right\rangle \left\langle \hat{a}\hat{b}^{\dagger }\right\rangle
+\left\langle \hat{a}\hat{b}\right\rangle \left\langle \hat{b}^{\dagger
2}\right\rangle \right) \right\}  \notag \\
&&\left. -\left\langle \hat{a}\hat{b}^{\dagger }\right\rangle \left(
3\left\langle \hat{b}^{\dagger 2}\right\rangle +1\right) \right] .
\end{eqnarray}%
\begin{eqnarray}
\frac{d}{dt}\left\langle \hat{a}^{\dagger }\hat{b}\right\rangle &=&i\left(
\Delta _{c}-\omega _{M}\right) \left\langle \hat{a}^{\dagger }\hat{b}%
\right\rangle +i\Omega \left\langle \hat{b}\right\rangle -\frac{1}{2}\left(
\Gamma _{a}+\Gamma _{b}\right) \left\langle \hat{a}^{\dagger }\hat{b}%
\right\rangle  \notag \\
&&+ig_{opt}\left[ \left\{ \left\langle \hat{a}^{\dagger }\hat{b}%
\right\rangle \left\langle \hat{b}^{\dagger 2}\right\rangle +2\left\langle 
\hat{a}^{\dagger }\hat{b}^{\dagger }\right\rangle \left\langle \hat{b}%
^{\dagger }\hat{b}\right\rangle \right\} -2\left\{ \left\langle \hat{a}%
^{\dagger 2}\right\rangle \left\langle \hat{a}\hat{b}^{\dagger
}\right\rangle +2\left\langle \hat{a}^{\dagger }\hat{a}\right\rangle
\left\langle \hat{a}^{\dagger }\hat{b}^{\dagger }\right\rangle \right\}
\right.  \notag \\
&&+2\left\{ \left( \left\langle \hat{a}^{\dagger }\hat{b}^{\dagger
}\right\rangle \left\langle \hat{b}^{2}\right\rangle +2\left\langle \hat{a}%
^{\dagger }\hat{b}\right\rangle \left\langle \hat{b}^{\dagger }\hat{b}%
\right\rangle \right) -\left( \left\langle \hat{a}^{\dagger 2}\right\rangle
\left\langle \hat{a}\hat{b}\right\rangle +2\left\langle \hat{a}^{\dagger }%
\hat{b}\right\rangle \left\langle \hat{a}^{\dagger }\hat{a}\right\rangle
\right) \right\}  \notag \\
&&\left. +\left\langle \hat{a}^{\dagger }\hat{b}\right\rangle \left(
3\left\langle \hat{b}^{2}\right\rangle +1\right) \right] .
\end{eqnarray}

\begin{eqnarray}
\frac{d}{dt}\left\langle \hat{a}\hat{b}\right\rangle &=&-i\left( \Delta
_{c}+\omega _{M}\right) \left\langle \hat{a}\hat{b}\right\rangle -i\Omega
\left\langle \hat{b}\right\rangle -\frac{1}{2}\left( \Gamma _{a}+\Gamma
_{b}\right) \left\langle \hat{a}\hat{b}\right\rangle  \notag \\
&&-ig_{opt}\left[ 2\left\langle \hat{a}\hat{b}^{\dagger }\right\rangle
+3\left\langle \hat{a}\hat{b}\right\rangle \left( \left\langle \hat{b}%
^{2}\right\rangle +1\right) \right] -ig_{opt}\left[ \left\langle \hat{a}\hat{%
b}\right\rangle \left\langle \hat{b}^{\dagger 2}\right\rangle +2\left\langle 
\hat{a}\hat{b}^{\dagger }\right\rangle \left\langle \hat{b}^{\dagger }\hat{b}%
\right\rangle \right]  \notag \\
&&-2ig_{opt}\left[ \left( 2\left\langle \hat{a}^{\dagger }\hat{a}%
\right\rangle \left\langle \hat{a}\hat{b}^{\dagger }\right\rangle
+\left\langle \hat{a}^{\dagger }\hat{b}^{\dagger }\right\rangle \left\langle 
\hat{a}^{2}\right\rangle \right) +\left( \left\langle \hat{a}\hat{b}%
^{\dagger }\right\rangle \left\langle \hat{b}^{2}\right\rangle
+2\left\langle \hat{a}\hat{b}\right\rangle \left\langle \hat{b}^{\dagger }%
\hat{b}\right\rangle \right) \right.  \notag \\
&&\left. +\left( 2\left\langle \hat{a}^{\dagger }\hat{a}\right\rangle
\left\langle \hat{a}\hat{b}\right\rangle +\left\langle \hat{a}^{\dagger }%
\hat{b}\right\rangle \left\langle \hat{a}^{2}\right\rangle \right) \right] .
\end{eqnarray}%
\begin{eqnarray}
\frac{d}{dt}\left\langle \hat{a}^{\dagger }\hat{b}^{\dagger }\right\rangle
&=&i\left( \Delta _{c}+\omega _{M}\right) \left\langle \hat{a}^{\dagger }%
\hat{b}^{\dagger }\right\rangle +i\Omega \left\langle \hat{b}^{\dagger
}\right\rangle -\frac{1}{2}\left( \Gamma _{a}+\Gamma _{b}\right)
\left\langle \hat{a}^{\dagger }\hat{b}^{\dagger }\right\rangle  \notag \\
&&+ig_{opt}\left[ 2\left\langle \hat{a}^{\dagger }\hat{b}\right\rangle
+3\left\langle \hat{a}^{\dagger }\hat{b}^{\dagger }\right\rangle \left(
\left\langle \hat{b}^{\dagger 2}\right\rangle +1\right) \right] +ig_{opt}%
\left[ \left\langle \hat{a}^{\dagger }\hat{b}^{\dagger }\right\rangle
\left\langle \hat{b}^{2}\right\rangle +2\left\langle \hat{a}^{\dagger }\hat{b%
}\right\rangle \left\langle \hat{b}^{\dagger }\hat{b}\right\rangle \right] 
\notag \\
&&+2ig_{opt}\left[ \left( \left\langle \hat{a}^{\dagger 2}\right\rangle
\left\langle \hat{a}\hat{b}\right\rangle +2\left\langle \hat{a}^{\dagger }%
\hat{a}\right\rangle \left\langle \hat{a}^{\dagger }\hat{b}\right\rangle
\right) +\left( \left\langle \hat{b}^{\dagger 2}\right\rangle \left\langle 
\hat{a}^{\dagger }\hat{b}\right\rangle +2\left\langle \hat{a}^{\dagger }\hat{%
b}^{\dagger }\right\rangle \left\langle \hat{b}^{\dagger }\hat{b}%
\right\rangle \right) \right.  \notag \\
&&\left. +\left( 2\left\langle \hat{a}^{\dagger }\hat{a}\right\rangle
\left\langle \hat{a}^{\dagger }\hat{b}^{\dagger }\right\rangle +\left\langle 
\hat{a}\hat{b}^{\dagger }\right\rangle \left\langle \hat{a}^{\dagger
2}\right\rangle \right) \right] .
\end{eqnarray}%
\begin{eqnarray}
\frac{d}{dt}\left\langle \hat{a}^{2}\right\rangle &=&-2i\Delta
_{c}\left\langle \hat{a}^{2}\right\rangle -2i\Omega \left\langle \hat{a}%
\right\rangle -\Gamma _{a}\left\langle \hat{a}^{2}\right\rangle
-2ig_{opt}\left\langle \hat{a}^{2}\right\rangle \\
&&-2ig_{opt}\left[ \left\langle \hat{a}^{2}\right\rangle \left( \left\langle 
\hat{b}^{\dagger 2}\right\rangle +\left\langle \hat{b}^{2}\right\rangle
+2\left\langle \hat{b}^{\dagger }\hat{b}\right\rangle \right) +2\left(
\left\langle \hat{a}\hat{b}^{\dagger }\right\rangle ^{2}+\left\langle \hat{a}%
\hat{b}\right\rangle ^{2}+2\left\langle \hat{a}\hat{b}\right\rangle
\left\langle \hat{a}\hat{b}^{\dagger }\right\rangle \right) \right] .  \notag
\end{eqnarray}%
\begin{eqnarray}
\frac{d}{dt}\left\langle \hat{a}^{\dagger 2}\right\rangle &=&2i\Delta
_{c}\left\langle \hat{a}^{\dagger 2}\right\rangle +2i\Omega \left\langle 
\hat{a}^{\dagger }\right\rangle -\Gamma _{a}\left\langle \hat{a}^{\dagger
2}\right\rangle +2ig_{opt}\left\langle \hat{a}^{\dagger 2}\right\rangle \\
&&+2ig_{opt}\left[ \left\langle \hat{a}^{\dagger 2}\right\rangle \left(
\left\langle \hat{b}^{\dagger 2}\right\rangle +\left\langle \hat{b}%
^{2}\right\rangle +2\left\langle \hat{b}^{\dagger }\hat{b}\right\rangle
\right) +2\left( \left\langle \hat{a}^{\dagger }\hat{b}^{\dagger
}\right\rangle ^{2}+\left\langle \hat{a}^{\dagger }\hat{b}\right\rangle
^{2}+2\left\langle \hat{a}^{\dagger }\hat{b}^{\dagger }\right\rangle
\left\langle \hat{a}^{\dagger }\hat{b}\right\rangle \right) \right] .  \notag
\end{eqnarray}%
\begin{eqnarray}
\frac{d}{dt}\left\langle \hat{b}^{2}\right\rangle &=&-2i\omega
_{M}\left\langle \hat{b}^{2}\right\rangle -\Gamma _{b}\left\langle \hat{b}%
^{2}\right\rangle -2ig_{opt}\left\langle \hat{a}^{\dagger }\hat{a}%
\right\rangle \\
&&-4ig_{opt}\left[ \left\langle \hat{a}^{\dagger }\hat{a}\right\rangle
\left( \left\langle \hat{b}^{2}\right\rangle +\left\langle \hat{b}^{\dagger }%
\hat{b}\right\rangle \right) +\left\langle \hat{a}\hat{b}\right\rangle
\left( \left\langle \hat{a}^{\dagger }\hat{b}^{\dagger }\right\rangle
+2\left\langle \hat{a}^{\dagger }\hat{b}\right\rangle \right) +\left\langle 
\hat{a}^{\dagger }\hat{b}\right\rangle \left\langle \hat{a}\hat{b}^{\dagger
}\right\rangle \right] .  \notag
\end{eqnarray}%
\begin{eqnarray}
\frac{d}{dt}\left\langle \hat{b}^{\dagger 2}\right\rangle &=&2i\omega
_{M}\left\langle \hat{b}^{\dagger 2}\right\rangle -\Gamma _{b}\left\langle 
\hat{b}^{\dagger 2}\right\rangle +2ig_{opt}\left\langle \hat{a}^{\dagger }%
\hat{a}\right\rangle \\
&&+4ig_{opt}\left[ \left\langle \hat{a}^{\dagger }\hat{a}\right\rangle
\left( \left\langle \hat{b}^{\dagger 2}\right\rangle +\left\langle \hat{b}%
^{\dagger }\hat{b}\right\rangle \right) +\left\langle \hat{a}^{\dagger }\hat{%
b}^{\dagger }\right\rangle \left( \left\langle \hat{a}\hat{b}\right\rangle
+2\left\langle \hat{a}\hat{b}^{\dagger }\right\rangle \right) +\left\langle 
\hat{a}^{\dagger }\hat{b}\right\rangle \left\langle \hat{a}\hat{b}^{\dagger
}\right\rangle \right] .  \notag
\end{eqnarray}


\begin{thebibliography}{99}
\bibitem{Haroche} S. Haroche and J.M. Raimond,\textquotedblleft Radiative
Properties of Rydberg States in Resonant Cavities\textquotedblright ,
Advances in Atomic and Molecular Physics \textbf{20}, 347 (1985).

\bibitem{Walther} D. Meschede, H. Walther and G. Muller,\textquotedblleft
One-Atom Maser\textquotedblright , Phys. Rev. Lett. \textbf{54}, 551 (1985).

\bibitem{Jaynes} E.T. Jaynes and F.W. Cummings,\textquotedblleft Comparison
of quantum and semiclassical radiation theories with application to the beam
maser\textquotedblright , Proc. IEEE \textbf{51,} 89 (1963).

\bibitem{Yablonovitch} E. Yablonovitch,\textquotedblleft Inhibited
Spontaneous Emission in Solid-State Physics and
Electronics\textquotedblright , Phys. Rev. Lett. \textbf{58}, 2059 (1987).

\bibitem{Gerard} J.M. Gerard, D. Barrier, J.Y. Marzin, R. Kuszelewicz, L.
Manin, E. Costard, V. Thierry- Mieg, and T. Rivera, \textquotedblleft
Quantum boxes a active probes for photonic microstructures: The pillar
microcavity case\textquotedblright , Appl. Phys. Lett. \textbf{69}, 449
(1996).

\bibitem{Lin} C.C. Lin, M.-C. Wu, B.-W. Shiau, Y.-H. Chen, I.A. Yu, Y.-F.
Chen, and Y.-C. Chen, \textquotedblleft Enhanced all-optical switching with
double slow light pulses\textquotedblright , Phys. Rev. A \textbf{86},
063836 (2012).

\bibitem{Hwang} J. Hwang, M. Pototschnig, R. Lettow, G. Zumofen, A. Renn, S.
Gotzinger, and V. Sandoghdar, \textquotedblleft A single-molecule optical
transistor\textquotedblright , Nature \textbf{460}, 76 (2009).

\bibitem{Dudin} Y.O. Dudin and A. Kuzmich, \textquotedblleft Strongly
Interacting Rydberg Excitations of a Cold Atomic Gas\textquotedblright ,
Science \textbf{336}, 887 (2012).

\bibitem{Saha} V. Venkataraman, K. Saha, P. Londereo and A.L. Gaeta,
\textquotedblleft Few-Photon all-Optical Modulation in a Photonic Band-Gap
Fiber\textquotedblright , Phys. Rev. Lett. \textbf{107}, 193902 (2011).

\bibitem{Kolchin} P. Kolchin, R.F. Oulton, and X. Zhang, \textquotedblleft
Nonlinear Quantum Optics ina Waveguide: Distinct Single Photons Strongly
Interacting at the Single Atom Level\textquotedblright , Phys. Rev. Lett. 
\textbf{106}, 113601 (2011).

\bibitem{Kippenberg} T.J. Kippenberg and K.J. Vahala, \textquotedblleft
Cavity Optomechanics: Back- Action at the Mesoscale\textquotedblright ,
Science \textbf{321}, 1172 (2008).

\bibitem{Grob} S. Gr\"{o}blacher, K. Hammerer, M.R. Vanner, M. Aspelmeyer,
\textquotedblleft Observation of strong coupling between a micromechanical
resonator and an optical cavity field\textquotedblright , Nature \textbf{460}%
, 724 (2009).

\bibitem{Wilson} I. Wilson-Rae, N. Nooshi, \ W. Zwerger, \ T.J.
Kippenberg,\textquotedblleft Theory of Ground State Cooling of a Mechanical
Oscillator Using Dynamical Backaction\textquotedblright , Phys. Rev. Lett. 
\textbf{99}, 093901 (2007).

\bibitem{Thomp} J.D. Thompson, B.M. Zwickl, A.M. Jayich, F. Marquardt, S.M.
Girvin, \textquotedblleft Strong dispersive coupling of a high-finess cavity
to a micromechanical membrane\textquotedblright , Nature \textbf{452}, 72
(2008).

\bibitem{Mas} M. Aspelmeyer, S. Gr\"{o}blacher, K. Hammerer, N. Kiesel,
\textquotedblleft Quantum optomechanics - throwing a
glance\textquotedblright , JOSA B, \textbf{27}, A189-A197 (2010).

\bibitem{Verhagen} E. Verhagen, S. Deleglise, S. Weis, A. Schliesser, and T.
J.Kippenberg, \textquotedblleft Quantum-coherent coupling of a mechanical
oscillator to an optical cavity mode\textquotedblright , Nature \textbf{482}%
, 63 (2012).

\bibitem{Chan} J. Chan, T. P. Mayer Alegre, A. H. Safavi-Naeini, J.T. Hill,
A. Krause, S. Gr\"{o}blacher, M. Aspelmeyer, \textquotedblleft Laser cooling
of a nanomechanical oscillator into its quantum ground
state\textquotedblright , Nature \textbf{478}, 89 (2011).

\bibitem{Bose} S. Bose, K. Jacobs and P. L. Knight,\textquotedblleft Scheme
to probe the decoherence of a macroscopic object\textquotedblright , Phys.
Rev. A \textbf{59}, 3204 (1999).

\bibitem{Santamore} D.H. Santamore, A.C. Doherty and M.C.
Cross,\textquotedblleft Quantum nondemolition measurement of Fock states of
mesoscopic mechanical oscillators\textquotedblright , Phys. Rev. B \textbf{70%
}, 144301 (2004).

\bibitem{Hohberger} C. Hohberger and K. Karrai,\textquotedblleft Cavity
cooling of a microlever\textquotedblright , Nature \textbf{432}, 1002 (2004).

\bibitem{Arcizet} O. Arcizet, P.F. Cohadon, T. Briant, M. Pinard, and A.
Heidmann, \textquotedblleft Radiation- pressure cooling and optomechanical
instability of a micromirror\textquotedblright , Nature \textbf{444}, 71
(2006).

\bibitem{Martin} I. Martin, and W.H. Zurek,\textquotedblleft Measureent of
Energy Eigenstates by a Slow Detector\textquotedblright , Phys. Rev. Lett. 
\textbf{98}, 120401 (2007).

\bibitem{Jacobs} K. Jacobs, P. Lougovski, and M. Blencowe,\textquotedblleft
Continous Measurement of the Energy Eigenstates of a Nanomechanical
Resonator without a Nondemolition Probe\textquotedblright , Phys. Rev. Lett. 
\textbf{98}, 147201 (2007).

\bibitem{Zwickl} J.D. Thompson, B.M. Zwckl, A.M. Jayich, F. Marquardt, S.M.
Girvin, and J.G.E. Harris, \textquotedblleft Strong dispersive coupling of a
high-finess cavity to a micromechanical membrane\textquotedblright , Nature
452, 72 (2008).

\bibitem{Zwickll} J.C. Sankay,C. Yang, B.M. Zwickl, A.M. Jayich, and J.G.E.
Harris, \textquotedblleft Strong and tunable nonlinear optomechanical
coupling in a low-loss system \textquotedblright , Nat. Phys. 6, 707 (2010).

\bibitem{Nori} Jie-Qiao Liao and F. Nori,\textquotedblleft Photon blockade
in quadratically coupled optomechanical systems \textquotedblright , Phys.
Rev. A \textbf{88}, 023853 (2013).

\bibitem{Scully} M.O. Scully and M. Suhail Zubairy, \textquotedblleft
Quantum Optics\textquotedblright , \ Cambridge University Press 1997.

\bibitem{Raymond} C.H. Raymond Ooi, Q. Sun, M. Suhail Zubairy,and M.O.
Scully, \textquotedblleft Correlation of photon pairs from the double Raman
amplifier: generalized analytical quantum Langevin Theory\textquotedblright
, Phys. Rev. A \textbf{75}, 013820 (2007).

\bibitem{Singh} S.K. Singh and C.H. Raymond Ooi, \textquotedblleft Quantum
correlations of quadratic optomechanical oscillator\textquotedblright , JOSA
B, \textbf{31}, 2390-2398 (2014).

\bibitem{Anglin} J.R. Anglin and A. Vardi, \textquotedblleft Dynamics of a
two-mode Bose-Einstein condensate beyond mean-field
approximation\textquotedblright , Phys. Rev. A \textbf{64}, 013605 (2001).
\end{thebibliography}
\end{document}